\documentclass[prl,twocolumn,showpacs,superscriptaddress,nofootinbib]{revtex4-2}

\pdfoutput=1
\usepackage{graphicx}
\usepackage{epstopdf} 
\usepackage{amsmath}
\usepackage{amssymb}
\usepackage{mathrsfs}
\usepackage{bm}
\usepackage{latexsym}
\usepackage{color}
\usepackage[usenames,dvipsnames]{xcolor} 
\usepackage[colorlinks=true,citecolor=MidnightBlue,linkcolor=RubineRed,urlcolor=MidnightBlue]{hyperref} 
\usepackage{float} 
\usepackage{multirow}         

\def\x{{\rm\bf x}}

\def\l{{\it  l}}

\newcommand{\beq}{\begin{equation}}
\newcommand{\eeq}{\end{equation}}
\newcommand{\beqa}{\begin{eqnarray}}
\newcommand{\eeqa}{\end{eqnarray}}

\newcommand{\tabincell}[2]{\begin{tabular}{@{}#1@{}}#2\end{tabular}}

\usepackage{tikz,xcolor,hyperref}
\definecolor{lime}{HTML}{A6CE39}
\DeclareRobustCommand{\orcidicon}{
\begin{tikzpicture}
\draw[lime, fill=lime] (0,0)
circle[radius=0.16]
node[white]{{\fontfamily{qag}\selectfont \tiny \.{I}D}}; 
\end{tikzpicture}
\hspace{-2mm}
}
\foreach \x in {A, ..., Z}{%
\expandafter\xdef\csname orcid\x\endcsname{\noexpand\href{https://orcid.org/\csname orcidauthor\x\endcsname}{\noexpand\orcidicon}}
}

\begin{document}
\title{Dissipation and Decay of Three Dimensional Holographic Quantum Turbulence}

\author{Hua-Bi Zeng\orcidB{}
}\email{zenghuabi@hainanu.edu.cn}
\affiliation{Center for Theoretical Physics , Hainan University, Haikou 570228, China}
\affiliation{Center for Gravitation and Cosmology, College of Physical Science and Technology, Yangzhou University, Yangzhou 225009, China}

\author{Chuan-Yin Xia}
\affiliation{Center for Gravitation and Cosmology, College of Physical Science and Technology, Yangzhou University, Yangzhou 225009, China}
\affiliation{Center for Theoretical Physics , Hainan University, Haikou 570228, China}

\author{Wei-Can Yang\orcidA{}
}
\affiliation{Department of Physics, Osaka Metropolitan University, 3-3-138 Sugimoto, 558-8585 Osaka, Japan}

\author{Yu Tian}\email{ytian@ucas.ac.cn}
\affiliation{School of Physical Sciences, University of Chinese Academy of Sciences, Beijing 100049, China $\&$ Institute of Theoretical Physics, Chinese Academy of Sciences, Beijing 100190, China}

\author{Makoto Tsubota}\email{tsubota@omu.ac.jp}
\affiliation{Department of Physics, Osaka Metropolitan University, 3-3-138 Sugimoto, 558-8585 Osaka, Japan}
\affiliation{Nambu Yoichiro Institute of Theoretical and Experimental Physics (NITEP), Osaka Metropolitan University, 3-3-138 Sugimoto, Sumiyoshi-ku, Osaka 558-8585, Japan}

\begin{abstract} 
{Quantum turbulence is a far-from-equilibrium process characterized by high nonlinearity. Holographic duality provides a systematic framework for simulating
the decaying $(3+1)$-dimensional quantum turbulence by numerically solving the dual Abelian-Higgs theory in a $(4+1)$-dimensional black hole background.
We reveal that different types of decay behavior of the total vortex line density $L$ emerge depending on the initial vortex line density, ranging
from $L\sim t^{-1.5}$ to $L\sim t^{-1}$, similar to the experimental observation of $^3$He in Phys. Rev. Lett. 96, 035301 (2006), and of $^4$He in Phys. Rev. Lett. 82, 4831 (1999)  and in Phys. Rev. Lett.  118, 134501 (2017).
Furthermore, by measuring the energy flux at the black hole horizon, we determine that the energy dissipation rate $dE/dt$ is proportional to the square of the total vortex line density, consistent with the vortex line decay equation proposed by W. F. Vinen and also the experimental measurement in Nature Physics 7, 473–476 (2011).}

\end{abstract}

\maketitle

\begin{figure}[t]
\centering
\includegraphics[trim=1cm 16cm 0cm 4cm, clip=true, scale=0.6, angle=0]{fig1.png}
\caption{{$3+1$ dimensional superfluid turbulence living on the boundary of an $AdS_5$ black hole,}
the energy dissipated in the decay dynamics will be absorbed by the black hole through its horizon.}\label{Fig1}
\end{figure}

Consider a superfluid, such as those realized in $^{3}$He and $^{4}$He, and stir it vigorously. This action creates a non-equilibrium state similar to that found in turbulent classical fluids, characterized by vortices occurring over a wide range of length scales. This phenomenon, known as ``quantum turbulence," has been the focus of extensive research since the 1980s \cite{Donnelly1986,Vinen2002,Vinen2008,Vinen2010,Nemirovskii2012,Barenghi2014,Barenghi_Skrbek_Sreenivasan_2023,TSATSOS20161}. ``Quantum" refers here to the defining property of superfluids where circulation is governed by topological quantization: vorticity can only occur in discrete amounts determined by the quantized circulation.  
A substantial body of experimental work is supported by simulations of effective phenomenological models such as the time-dependent Ginzburg-Landau (Gross-Pitaevski, G-P \cite{Gross1961,Pitaevskii1961,Pitaevskii1958}) model and ``vortex filament model" (VFM) \cite{Schwarz1985,Saffman1992,Hanninen2014}. However, there is still a lack of complete consistency between experiments, theory, and numerical simulations in understanding the decay of quantum turbulence, as comprehensively reviewed in \cite{Tsubota2017,Barenghi_Skrbek_Sreenivasan_2023}. 

\begin{figure*}[t]
\centering
\includegraphics[scale=1, angle=0]{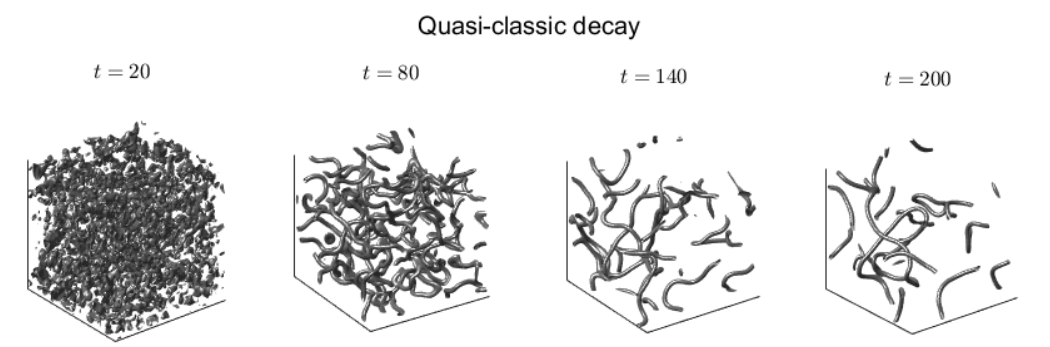}
\includegraphics[scale=1, angle=0]{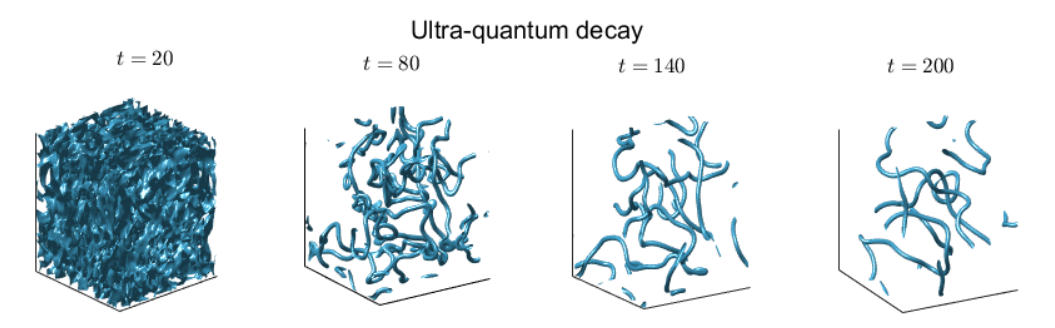}
\caption{{Isosurfaces of the superfluid density } for the turbulent flows on the $3+1$ dimensional boundary of the $4+1$ dimensional black hole spacetime.
Evolution of vortex lines for two initial vortex line densities, top row: $L \propto t^{-1.5}$ decay, bottom row: $L \propto t^{-1}$ decay. Except the density  no other fundamental difference between the two kinds of decay can be seen.
}\label{Fig2}
\end{figure*}

{Quantum turbulence encompasses two distinct types: quasi-classical (Kolmogorov) turbulence and ultra-quantum (Vinen) turbulence,
which are characterized by different features in the decay of quantized vortices due to energy dissipation.
Superfluid turbulence experiments involving both $^4$He \cite{Stalp1999,Walmsley2008,Walmsley2007,stalp2002dissipation,Walmsley2017,Gao2016} and
$^3$He \cite{Bradley2006,Bradley2011}
demonstrate decay dynamics that the decay of the {vortex line density (length of the vortex line per
unit volume)} $L$ conforms to the scaling law
\begin{equation}
L(t) \propto t^{-3/2},\label{eq1}
\end{equation}
which is called quasi-classical turbulence.}
The other type, ultra-quantum turbulence, admits 
\begin{equation}
L(t) \propto t^{-1},\label{eq2}
\end{equation}
which have also been observed in both $^4$He \cite{Walmsley2017,Gao2016,Gao} and $^3$He \cite{Bradley2006} when the temperature is low enough. The primary difference
between the two types depends on whether the dominant
dynamics occurs at scales above or below the
mean intervortex distance $\ell \sim L^{-1/2}$, which corresponds to dense vortex density and dilute vortex density, respectively.
In the quasi-classical case when the flow occurs on a scale greater than $\ell$, the emergence of large-scaled quasi-classical vortices arises from the correlations in vortex line polarization \cite{volovik2003classical}, so the energy spectrum should follow the Kolmogorov scaling as the classical turbulence, as confirmed by both G-P \cite{Kerr2011,Makoto2006} and VFM simulation \cite{Baggaley2012}.
{On the contrary}, in the ultra-quantum case, the resulting uncorrelated entanglement has no classical correspondence and therefore exhibits completely different dynamics \cite{1957RSPSA.240..128V,1957RSPSA.240..114V,1957RSPSA.242..493V,1958RSPSA.243..400V}.
The results show that both the G-P \cite{Makoto2006} and VFM simulations \cite{Vinen2005, Vinen2003} reveal a $t^{-3/2}$  decay as well as a $t^{-1}$ decay. 
Some enlightenment can be observed from the simulation that quasi-classical decay of $-3/2$ is obtained for large density vortex tangle, but ultra-quantum decay of $-1$ is obtained for dilute vortex with vortex reconnection.

In both cases, {it was first proposed by Vinen \cite{PhysRevB.61.1410} that} the dissipation rate of flow energy $E$ should take the same form
\begin{equation}
\frac{dE}{dt}\propto -\nu L^2. \label{eq3}
\end{equation}
Here $\nu$ is the ``effective kinematic viscosity'' {which can be 
measured in experiments \cite{Walmsley2008,Walmsley2007}.}
When the density is dilute, it is natural to expect that $E \propto L$ \cite{Vinen2010}, thus obtaining a pure quantum decay $L \propto t^{-1}$. However, in the quasi-classical case, if taking the classical energy decay behavior $\frac{dE}{dt} \propto -t^{-3}$ and substituting it into the formula (\ref{eq3}), then $L \propto t^{-1.5}$ can be
obtained \cite{Vinen2010}.

{At zero temperature or very low temperatures, the dissipation (\ref{eq3}) is mainly due to emission of sound waves at large wave numbers. It is intriguing, as shown by experiments \cite{Stalp1999}, that the validity of Vinen's equation (\ref{eq3}) extends to the finite temperature case, where reliable theoretical arguments are lacking. For numerical simulation with effective models, dissipation at a finite temperature is typically handled using phenomenological parameters, while a rigorous treatment of such dissipation is generally challenging as well.} In this Letter, we present results based on a rigorous, first-principles approach, albeit with certain restrictions. This unconventional approach specifically addresses the process of dissipation {at a finite temperature}. 
First of all, in this approach the system has a fully consistent thermodynamics for equilibrium states. Further, non-equilibrium superfluid dynamics with dissipation at a finite temperature is realized as irreversible dynamics in presence of a ``hairy" black hole. Ultimately, the relaxation of the turbulent ``vortex tangle" is translated into the manner in which energy is absorbed by this black hole when vortex tubes sweep over its horizon (Fig. \ref{Fig1}).

This refers to a mathematical contraption that originates in string theory: the AdS/CFT correspondence  \cite{Maldacena1998,Gubser1998,Witten1998} , which is a holographic duality map that relates the physical properties of a material system in D (``boundary") space time dimensions to a gravitational (general relativity) problem in D+1 dimensions (``bulk")  \cite{Ammon2015,Zaanen:2015oix,Jan2021,Liu2018,Hartnoll2009}. {Remarkably, it was discovered that the universal properties of superfluid states can be described by a $U(1)$ symmetry broken theory living in an $AdS$ black hole background \cite{Gubser2008,Hartnoll2008,2009holosuperfluid}.} A special benefit is that this also captures non-equilibrium dynamics in terms of a dynamical gravitational evolution yielding a first-principles framework also of dissipative aspects at a finite temperature, resulting in a non-perturbative effective description at
strong coupling.

Within certain restrictions it is also possible to  numerically simulate the physics when many vortices are present in a two spatial dimensional superfluid \cite{Hong2014,2023PhRvL.131v1602L,2021PhRvL.127j1601W,Lan:2016cgl,2023PhRvB.107n4511Y,2023PhRvB.107n4511Y,Yang2024hom}. An early success is the demonstration of the direct cascade referring to the flow of energy from larger to smaller scales in a quantum-turbulent fluid in two space dimensions \cite{Hong2014,du2015}. Here we will take a step further by addressing quantum turbulence in three space dimensions with bulk action
\begin{equation}\label{action}
S=\int d^5x \sqrt{-g}\Big[R+ \frac{\Lambda}{L^2}+\frac{1}{q^2}\big(-\frac{1}{4}F^2-( |D\Psi|^2-m^2|\Psi|^2)\big)\Big],
\end{equation}
{where $F^2:=F^{MN}F_{MN}$, the electromagnetic field $F_{MN}=\partial_M A_N-\partial_N A_M$, and $D_M:=\partial_M-i A_M$. In the so-called probe limit we ignore the backreaction of the matter fields onto the geometry. Solving only the gravity part of the action (\ref{action}) yields the $AdS_5$ Schwarzschild black hole background geometry
\begin{equation}
ds^2=\frac{\l^2}{u^2}(-f(u) dt^2 -2dtdu+ dx^2+dy^2+dz^2),
\end{equation}
where $f(u)=1-(u/u_h)^4$, $u$ is the extra bulk dimension. According to AdS/CFT correspondence, $u$ roughly
corresponds to the RG scale of the dual field theory, interpolating
between IR physics near the horizon ($u=1$) and UV physics near the boundary ($u=0$).
The temperature of the $3+1$ dimensional superfluid system is set by the Hawking temperature $T = (\pi u_h)
^{-1}$. When $T$ goes below a critical value $T_c$, a cloud of the complex scalar $\Psi$
builds up in the bulk that spontaneously breaks the U(1) symmetry, corresponding to a second-order phase transition of the boundary system into the superfluid state with a non-vanishing condensate.}
The details of the model are given in \cite{Supplementary}. 
{This setup in the probe limit} is mathematically consistent at temperatures that are not too low compared to the superfluid $T_c$ \cite{HHH2008} where the normal fluid density is quite high and so still does not contribute to turbulence, just acting as a heat bath to  dissipate the energy of the vortex system \cite{Hong2014}.
But these are precisely the conditions governing the quantum turbulence in superfluid $^3$He away from the very low temperature regime \cite{Vinen2010}, where the normal fluid does not contribute to turbulence due to its large viscosity.

\begin{figure}[t]
\centering
\includegraphics[trim=0.5cm 0cm 0cm 0cm, clip=true,scale=0.65, angle=0]{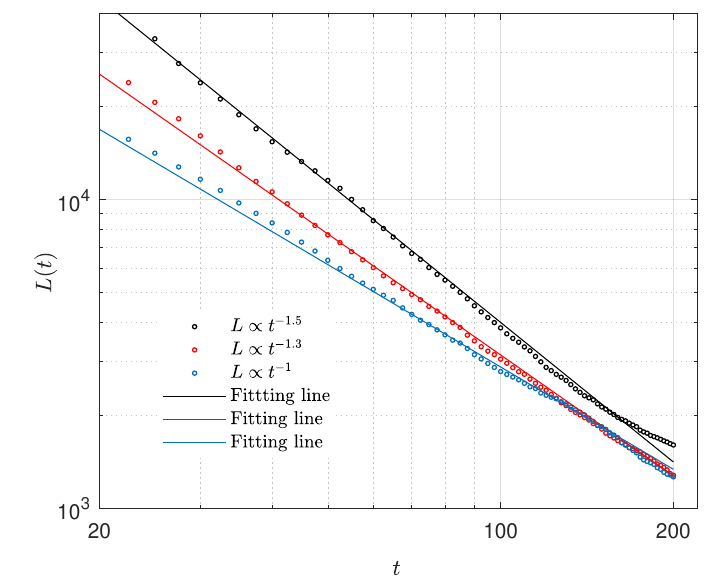}
\includegraphics[trim=0.5cm 0cm 0cm 0cm, clip=true,scale=0.65, angle=0]{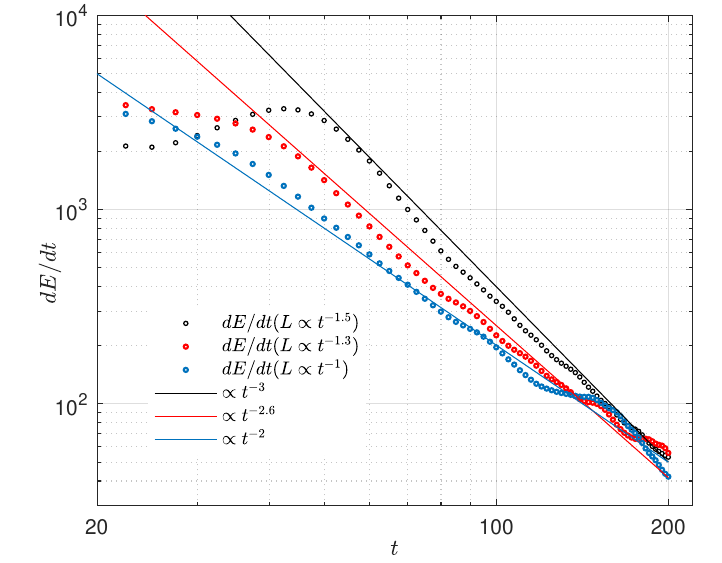}
\caption{Total vortex lines length decay behavior and it's energy dissipation rate
though the computation of the energy flux through the horizon.}\label{Fig3}
\end{figure}

We now describe three-dimensional superfluid turbulent flow dynamics in holography by numerically solving
the bulk equations of motion. We set $T=0.83T_c$ and work in a $50\times50\times50$ periodic box,
with $181$ Fourier points in every direction. 
The coherent length of a single vortex is therefore $\xi \sim 3.62$. This means that we describe a single vortex with a $20\times20$
grid, which is fully adequate.
The way to generate a turbulence is to use an initial
uniform {superfluid} state plus $N$ randomly
distribute vortices of winding number $W= \pm 1$ on every $x-y, y-z,$ and $x-z$ slice/plane (see \cite{Supplementary} for numerical details).
This method is close to the method of generating turbulence in
superfluid $^3$He by oscillating grid \cite{Bradley2006}, which is the experiment that the
holographic simulation will be mainly compared to.
With the chaotic velocity field of the initial randomly distributed vortices,
the system will evolve according to the equations of motion.
The vortex lines will be developed very soon and can be observed obviously at $t \approx 20$.
By tuning the numbers $N$
of vortices we are able to generate different densities of vortex lines.
In Fig. \ref{Fig2} we show the dynamic evolution of turbulence for two case of initial vortex densities, the dense case with $N=100$ (black dots)
and the dilute case with $N=20$ (blue dots). In the upper panel of Fig. \ref{Fig3}, we present a log-log plot showing both the decay dynamics of the total vortex line density.  
From the simulation we find that the turbulence
exhibits two decay scaling: $L \propto t^{-1.5}$ in the dense case while $L \propto t^{-1}$ in the dilute case for
$20<t<200$. This behavior is consistent with observations in superfluid $^3$He-B at low temperature regime where the normal fluid density is negligible and turbulence is mainly induced by vibrating superfluid part \cite{Bradley2006}, that in the dense initial vortex line case the $t^{-3/2}$ decay was observed while for the dilute case the $t^{-1}$ decay appeared.
When $t>200$, there are only a
few vortex lines left, making it difficult to define the state as turbulent.
Due to the presence of the bulk black hole, energy dissipation is manifested as irreversible energy absorption by the horizon.
Consequently, the energy of the superfluid dissipates as a positive energy flux through the horizon \cite{Hong2014,Tian:2023},
which is defined as
\begin{equation}
\frac{d E}{d t}=-\int d^3 x \sqrt{-g} \mathcal T^u_t (t,{\bf x},u)|_{\rm horizon}
\end{equation}
with $\mathcal{T}_N^M$ the stress tensor of $\Psi$ and $A_M$ in the bulk
\begin{align}
\nonumber
\mathcal T^{M}_{\ N} = {\textstyle \frac{1}{2}} \big \{  F_{NA}F^{MA} - {\textstyle \frac{1}{4} } \delta^{M}_{\ N} F_{AB}F^{AB}
+ D_N \Phi^* D^M \Phi \\ \label{eq:bulkstress}
+ D^M \Phi^* D_N \Phi - {\textstyle \frac{1}{2}} \delta^{M}_{\ N}  \left ( D_A \Phi^* D^A \Phi + m^2 \Phi^* \Phi \right ) \big \}.
\end{align}
At the horizon, we have 
\begin{equation}
\mathcal T^u_t|_{u=1}=\frac{1}{2} ( F_{0i}F^{ui}  + D_0 \Psi^* D^u \Psi
+ D^u \Psi^* D_0 \Psi )|_{u=1}.
\end{equation}
A sample configuration of energy flux is given in \cite{Supplementary}, which is zero nearly everywhere except at the location of vortex lines.
In the lower panel of Fig. \ref{Fig3} we plot the energy dissipation rate measured by the energy flux cross the horizon,
the $t^{-1.5}$ decay and $t^{-1}$ decay correspond to scaling $dE/dt \propto t^{-3}$ and $dE/dt \propto t^{-2}$ respectively.
Importantly, the $dE/dt \propto t^{-3}$ energy decay behavior {is the same as in} the direct measurement of the energy dissipated by quantum turbulent $^3$He \cite{Bradley2011}
{in the quasi-classical regime at low temperatures, which confirms the universality of Eq.(\ref{eq3}) at different temperatures.} An interesting observation is that, if we chose $N=50$ (red dots), the decay follows $t^{-1.3}$ with corresponding energy dissipation rate $t^{-2.6}$.
This observation is exactly the prediction from the decay equation \ref{eq3}, which may can be understood as the following:
the energy dissipation rate per unit length $(dE/dt)/L$ is proportional to the the vortex line density $L$,
by assuming that the energy dissipation is mainly through the reconnection of vortex lines, whose rate is naturally 
proportional to the vortex line density. Then different
decay scaling of vortex lines can be understood as a result of how the total energy $E$ of vortex lines depends on $L$.
In the dilute case, where the interactions between 
lines can be ignored, a linear relationship
between $E$ and $L$ is expected. In the case with dense vortex lines, the total energy relationship differs.
From this perspective, we can also expect that all decays at later times, when the vortex lines are dilute, will approach the same $t^{-1}$ scaling at  $t \sim 170$, as shown in Fig. \ref{Fig3}. {More 
interestingly, the crossover from $t^{-1.5}$ decay to $t^{-1}$ decay
by varying the initial vortex line density realized in the holographic superfluid turbulence is very similar to the experimental observation in $^4$He (see Fig.2 in \cite{Walmsley2017}).}

Another characteristic of turbulence comes from the scaling of the energy spectrum.
In superfluid turbulence the  ``kinetic energy" for
the superfluid field $\langle\psi(t,\textbf{x})\rangle$ of the boundary field theory is defined
as
\begin{eqnarray}\label{energyspectrum}
E_{{k}} (k)=\frac{1}{2}\int^\infty_0 d\theta k^2 \mathcal{V}^{\star} ({\bf k}) \mathcal{V}({\bf k}),
\end{eqnarray}
where $\mathcal{V}=\langle \psi \rangle {\bf v}$ and ${\bf v}$ is the superfluid velocity $\frac{i}{2}[\langle \psi^{\star} \rangle \nabla \langle \psi \rangle - \langle \psi \rangle \nabla \langle \psi \rangle ^\star ] / |\langle \psi \rangle|^2$.

\begin{table*}[]
\tiny
\begin{tabular}{|l|l|l|l|l|l|l|l|l|l|l|l|l|l|}
\hline
\multirow{2}{*}{$ L(t)$} & \multirow{2}{*}{\quad Systems observed} & \multicolumn{4}{l|}{$\quad \quad \quad d E/ d t$} & \multicolumn{4}{l|}{$\quad \quad E(k)$}  \\ \cline{3-10}
     &          & EXP & G-P & VFM & Holo & EXP & G-P  & VFM & Holo  \\ \hline
$ t^{-3/2}$ &\tabincell{c}{ $^4 He$\cite{Stalp1999,Walmsley2008,Walmsley2007,stalp2002dissipation,Walmsley2017,Gao2016}, \\ $^3 He$\cite{Bradley2006,Bradley2011}} &  $t^{-3}$\cite{Bradley2011}         &  $t^{-3}$ \cite{Kerr2011,Makoto2006}  &  $t^{-3}$ \cite{Baggaley2012} & $t^{-3}$   & $k^{-5/3}$ \cite{Maurer1998,Salort2010}        & $k^{-5/3}$\cite{Kerr2011,Makoto2006}  &  $k^{-5/3}$ \cite{Baggaley2012,PhysRevLett.109.205304}   & $k^{-5/3}$  \\ \hline
$ t^{-1}$   & $^3 He$\cite{Bradley2006}, $^4 He$\cite{Walmsley2008,Walmsley2017} & UKN     &  $t^{-3}$\cite{Makoto2006}   &  UKN   & $t^{-2}$   & UKN          & \tabincell{c}{$k^{-5/3}$\cite{Makoto2006,Cidrim2017},\\ no $k^{-5/3}$ \cite{PhysRevA.94.053632},\\ $k^{-1}$\cite{Cidrim2017}}  &  $k^{-1}$\cite{Vinen2005,Vinen2003,Kozik2005,Baggaley2012,PhysRevLett.109.205304,araki2002energy,barenghi2016regimes}  & $k^{-5/3}$\\ \hline
\end{tabular}
\caption{{Decay of {vortex line density} and it's corresponding energy dissipation rate, energy spectrum from experiments (EXP), Gross-Pitaevskii (G-P) equation simulation, vortex filament model (VFM) simulation and holographic simulation.}
}\label{table}
\end{table*}

{An important fact of such a spectrum is that it often has a Kolmogorov scaling behavior in certain range of $k$, similar to classical turbulence.} The Kolmogorov spectrum $E_{k} \propto k^{-5/3}$ \cite{1941Kolmogorov} has been observed in $^4$He \cite{Maurer1998,Salort2010}, which may be
understood using the idea that the inviscid superfluid and the viscous normal fluid are likely to be coupled together by the mutual friction between them and thus to behave like a conventional fluid \cite{PhysRevB.61.1410}. 
Numerical simulation using the G-P equation confirmed the quantum turbulence purely from the superfluid part also shows the $k^{-5/3}$ law \cite{PhysRevLett.94.065302}, which may
support that the Richardson cascade process works in the system where the dissipation is
caused mainly by removing short wavelength excitations emitted at vortex reconnections.

\begin{figure}[t]
\includegraphics[ scale=1, angle=0]{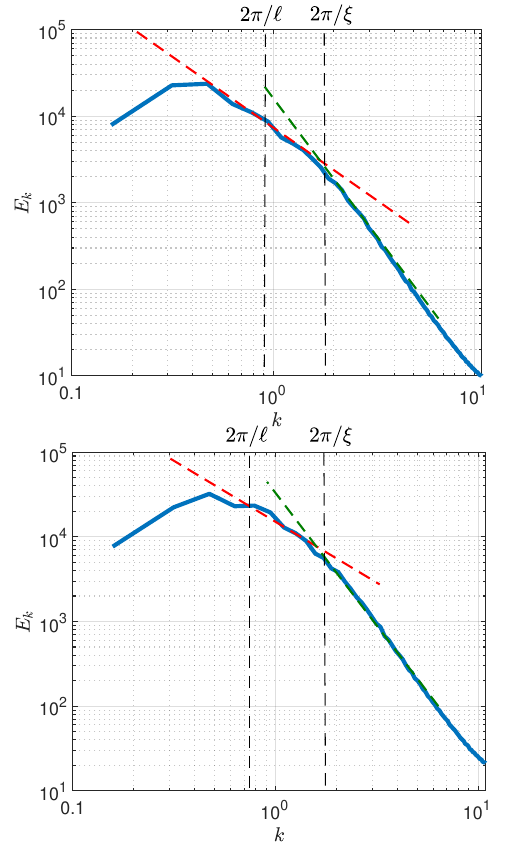}
\caption{Kinetic energy spectrum in the well defined turbulent region $t=80$ for classical decay (top) and quantum decay (bottom). The red line is the $k^{-5/3}$ line while the green line is the $k^{-3}$ line.}\label{Fig4}
\end{figure}

In Fig. \ref{Fig4} we plot the spectrum of our simulation in the well-defined turbulence region. For both the {quasi-classical and ultra-quantum} cases the $k^{-5/3}$ law
always shows up, but for the dilute case the $k^{-5/3}$ law is less evident. {The Kolmogorov spectrum observed in the quasi-classical (Kolmogorov)
turbulence is consistent with both G-P simulation and VFM simulation, which comes from bundles of coherent vortices \cite{Baggaley2012}.
But for the ultra-quantum (Vinen) turbulence the energy spectrum results obtained by the three methods are not quite consistent, as summarized in Table. \ref{table}. In the Vinen turbulence, most of
the energy is expected at wave number $2 \pi/\ell$ and there is no $k^{-5/3}$ scaling at large $k$, but a $k^{-1}$
spectrum should appear when the vortex lines are randomly oriented to each other (the spectrum of an
isolated vortex line) \cite{Baggaley2012,PhysRevLett.109.205304}. The holographic
simulation confirms the universality of Vinen's equation (\ref{eq3}), suggesting there may exist another kind of $t^{-1}$ decay turbulence 
with bundles of coherent vortex lines as long as the density is dilute.} {The $k^{-3}$ law in the ultraviolet regime ($k>2\pi/ \xi$) is confirmed to be related to the spectrum of the discrete vortex structure with the help of G-P equation simulations \cite{nore1997decaying,Bradley2012}. Although there is some difference for a single vortex configuration between
Gross-Pitaevskii-like and holographic superfluids \cite{2010PhRvDholovortex}, the superfluid velocity {and condensate
configuration near a quantum vortex core must have similar behaviors}, so the kinetic energy spectra in the large $k> 2\pi/\xi$ region should be the universal
on physical grounds \cite{-3spectrum2015,Hong2014}.}

An experimental about a quasi-classic decaying
quantum turbulence in superfluid $^{4}$He found that local velocity distribution can distinguish between quantum and classical turbulence \cite{PhysRevE.67.047302,PhysRevLett.101.154501Vd},
because quantum vortex reconnection in superfluid turbulence is a high speed
event admitting a statistic probability $P(v_x) \propto v^{-3}_x$ at large speed, different from the Gaussian velocity distribution in classical turbulence \cite{PhysRevLett.104.075301}.
In \cite{Supplementary},  we present the velocity distribution results in holographic quantum turbulence, both quasi-classical and unltra-quantum decay admit the same power law $P(v_x) \propto v^{-3}_x$ at large speed.

{In summary, with the advantage of the holographic method that the dissipation rate can be measured as the horizon energy flux, the study of tangled vortex line dynamics in 3D holographic quantum turbulence shows good agreement with the decay equation (\ref{eq3}) proposed by Vinen. We expect that Eq. (\ref{eq3}) is universal due to the physical understanding that the energy decay rate per unit vortex line length is proportional to the vortex line reconnection rate, which is also proportional to the vortex line density. This understanding allows us to explain the crossover observed in $^4$He experiment \cite{Walmsley2017} and the holographic simulation from the $t^{-1.5}$ decay to the
$t^{-1}$ decay more comprehensively, which is expected to be tested in future experiments.}

{We found that \cite{2024arXiv241022410W} has some overlap with the present work while this paper was in the review process.}

{\it Acknowledgements.}
We especially thank Jan Zaanen for his very 
valuable comments and advices at the early stage of the project. Y. T. thanks Yu-Kun Yan for very helpful discussions.
H.B. Z. acknowledges the support by the National Natural Science Foundation of China (under Grants No. 12275233). M. T. acknowledges the support by the
JSPS KAKENHI (under Grant No. JPK23K03305 and JPH22H05139). Y. T. acknowledges the support by the National Natural Science Foundation of China (under Grants No. 12035016, 12375058 and 12361141825).

\bibliography{3DholoTurb_arxiv}

\clearpage
\onecolumngrid
\begin{appendix}
\begin{center}

\renewcommand\thefigure{S\arabic{figure}}    
\setcounter{figure}{0} 
\renewcommand{\theequation}{S\arabic{equation}}
\setcounter{equation}{0}
\renewcommand{\thesubsection}{SI\arabic{subsection}}

\end{center}
\setcounter{equation}{0}
\setcounter{table}{0}

{\bf Supplementary materials for "Dissipation and Decay of Three-Dimensional  Holographic Quantum Turbulence"}

\section{$AdS_5$ black hole metric and bulk equations of motion}
\label{appendixA}
The model we used is the standard minimal holographic model of $U(1)$ symmetry broken physics \cite{Gubser2008,Hartnoll2008} but defined in the $AdS_5$ background instead of the original $AdS_4$.
The holographic theory is the  Abelian-Higgs-Einstein model of local $U(1)$ gauge field $\textbf{A}$ and charged
scalar field $\Psi$ coupled to an $AdS$ black hole. Under the standard AdS/CFT dictionary,
the conserved boundary current $J_M(x,u)$ is mapped to the dynamical $U(1)$
gauge field $A_M(x, u)$ in the gravitational bulk, while the scalar operator $\psi$ is mapped to
a bulk scalar field $ \Psi$.
In the unit $\hbar = c = G_N = 1$, the action of the theory is
\begin{equation}
S=\int d^5x \sqrt{-g}\Big[R+ \frac{\Lambda}{L^2}+\frac{1}{q^2}\big(-\frac{1}{4}F^2-( |D\Psi|^2-m^2|\Psi|^2)\big)\Big].
\end{equation}

\begin{figure}[t]
\centering
\includegraphics[scale=0.6, angle=0]{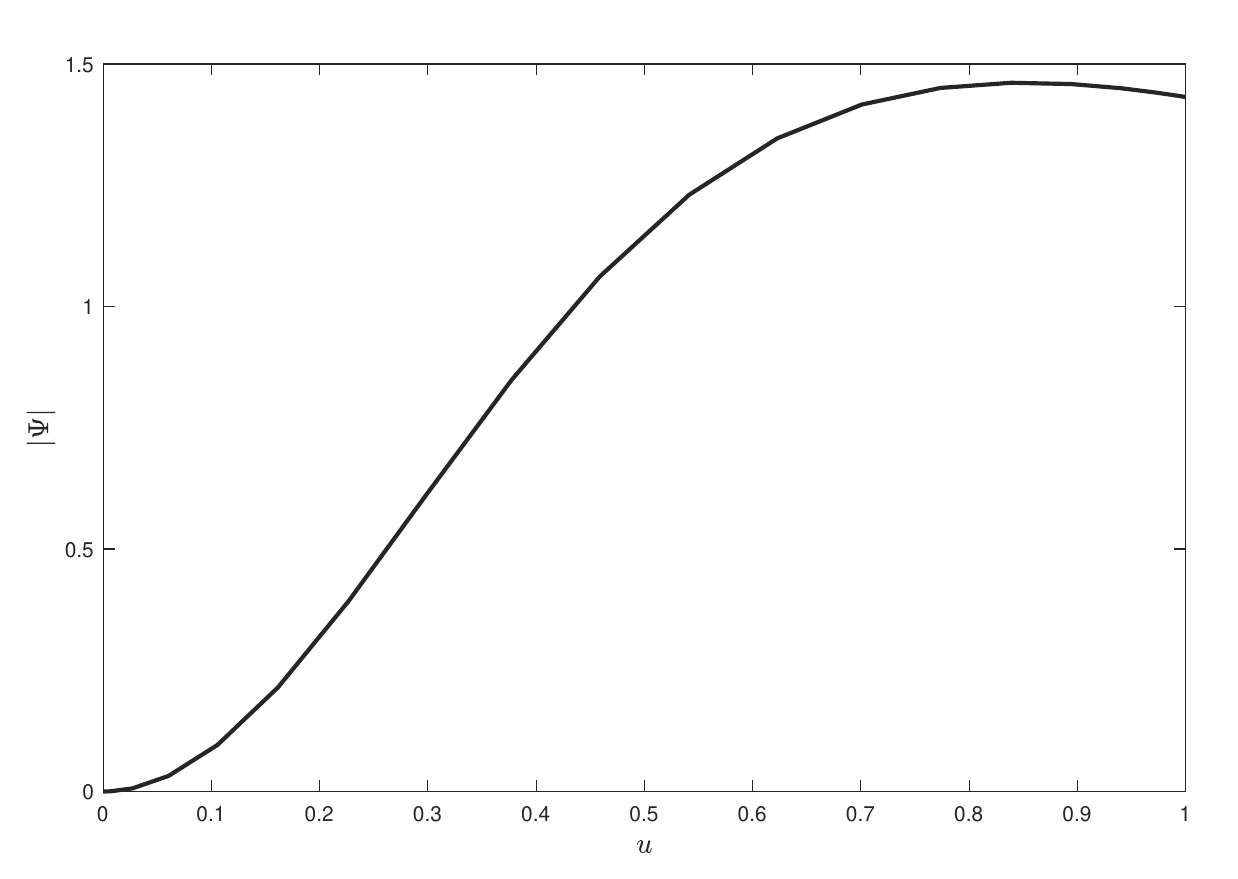}
\caption{{For a uniform superfluid state, the profile of
$|\Psi|$.}}\label{Figs1}
\end{figure}

If the Abelian-Higgs model with only quadratic potential of scalar field is defined in a
flat space time there is no symmetry broken, a quartic potential is needed. However, when the
charged scalar field coupled to  a negative cosmological
constant gravity, the scalar field will condensate (stable finite value solution) when the black hole
temperature is below a critical value \cite{Gubser2008}.
\begin{figure}[t]
\centering
\includegraphics[scale=0.6, angle=0]{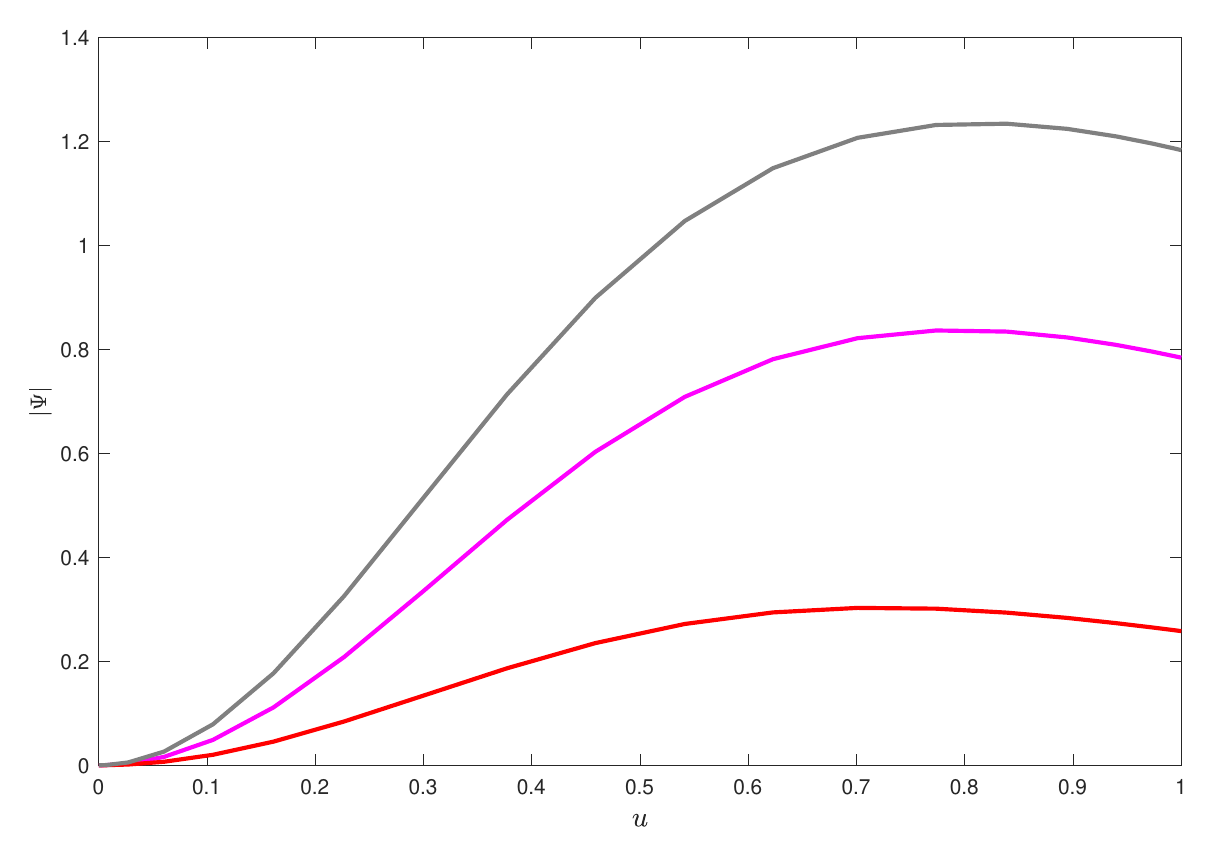}
\caption{{The profiles of
the $|\Psi|$ for different positions from away from a vortex line to a position close to a vortex line core, the distances to the vortex core are 9 (gray line), 5 (blue line) and 3 (red line).}}\label{Figs2}
\end{figure}

Following \cite{Hartnoll2008} we work in the probe limit, which applies when the charge $q$ of $\Psi$ is
large. In this limit the back-reaction from the matter fields is ignored,
then the gravitational system is approximated by an Abelian Higgs model
defined in a Schwarzschild black hole background geometry
\begin{equation}
ds^2=\frac{\l^2}{u^2}(-f(u) dt^2 -2dtdu+ dx^2+dy^2+dz^2),
\end{equation}
where $f(u)=1-(u/u_h)^4$, $u$ is the extra bulk dimension,
$u_h$ is the horizon while $u=0$ is the AdS boundary.
The black hole's Hawking temperature $T$ is proportional to the
$u_h$, and there is a critical $T_c$ below which the scalar field
will condense. Without loss of generality we can set $L=1$.

The equations of motion of the fields $A_M$ and $\Psi$ reads
\begin{equation}
d_{N} F^{M,N}=J^{N},\quad (-D^2+m^2) \Psi=0,
\end{equation}
work in the axial gauge $A_{u}=0$ to fix the gauge degree of freedom, we have highly nonlinear
coupled PDEs for the five fields $A_t, A_x, A_y, A_z, \Psi$.
The fully expanded equations of motion can be written
as
\begin{align}
\label{eomf1}
m^2\Psi+3u(i A_t\Psi+f\partial_u\Psi-\partial_t \Psi)+u^2 [\Psi (A_x^2+A_y^2+A_z^2+i(-\partial_u A_t+\partial_x A_x+\partial_y A_y+\partial_z A_z))+\nonumber\\
2i(- 
A_t\partial_u\Psi+A_x\partial_x\Psi+A_y\partial_y\Psi+A_z\partial_z\Psi)-\partial_x^2\Psi -\partial_y^2\Psi-\partial_z^2\Psi-\partial_u f \partial_u \Psi-f \partial_u^2 \Psi+2 \partial_t \partial_u \Psi]=0
\end{align}
\begin{align}
\label{eomf2}
u \partial_u A_t+u^2 \partial_u(-\partial_u A_t+ \partial_x A_x+ \partial_y A_y+ \partial_z A_z)+i \Psi^* \partial_u \Psi-i \Psi \partial_u \Psi^*=0
\end{align}
\begin{align}
\label{eomf3}
 2 A_x |\Psi|^2+i \Psi^* \partial_x \Psi-i \Psi \partial_x \Psi^*+u (\partial_x A_t-\partial_t A_x+f\partial_u A_x)-u^2[\partial_u(f\partial_u A_x) +\partial_y^2 A_x+\partial_z^2 A_x-\partial_x( \partial_y A_y+ \partial_z A_z)\nonumber\\
+\partial_u \partial_x A_t-2 \partial_t \partial_u A_x ]=0   
\end{align}
\begin{align}
\label{eomf4}
 2 A_y |\Psi|^2+i \Psi^* \partial_y \Psi-i \Psi \partial_y \Psi^*+u (\partial_y A_t-\partial_t A_y+f\partial_u A_y)-u^2[\partial_u(f\partial_u A_y) +\partial_z^2 A_y+\partial_x^2 A_y-\partial_y( \partial_z A_z+ \partial_x A_x)\nonumber\\
+\partial_u \partial_y A_t-2 \partial_t \partial_u A_y ]=0   
\end{align}
\begin{align}
\label{eomf5}
 2 A_z |\Psi|^2+i \Psi^* \partial_z \Psi-i \Psi \partial_z \Psi^*+u (\partial_z A_t-\partial_t A_z+f\partial_u A_z)-u^2[\partial_u(f\partial_u A_z) +\partial_x^2 A_z+\partial_y^2 A_z-\partial_z( \partial_x A_x+ \partial_y A_y)\nonumber\\
+\partial_u \partial_z A_t-2 \partial_t \partial_u A_z ]=0   
\end{align}
\begin{align}
\label{eomf6}
2 A_t |\Psi|^2 + i \Psi^* \partial_t \Psi-i \Psi \partial_t \Psi^*
+f (-i \Psi^* \partial_u \Psi+i \Psi \partial_u \Psi^*) - u^2 [\partial_x^2 A_t+\partial_y^2 A_t+\partial_z^2 A_t+f\partial_u(\partial_x A_x+\partial_y A_y+\partial_z A_z)\nonumber \\
-\partial_t(\partial_u A_t+\partial_x A_x+\partial_y A_y+\partial_z A_z)]=0
\end{align}
These six partial differential equations 
Eq. \eqref{eomf1}-Eq. \eqref{eomf6} are not independent, so we can choose any five of them. In this work, we have choose Eq. 
\eqref{eomf1}-Eq. \eqref{eomf5} while the rest Eq. \eqref{eomf6} can be used to check the self-consistency.

At the horizon, in our ingoing coordinates, physical solutions should be
regular. Near the boundary, a general solution takes the following form
\begin{equation}
A_\nu(t,{\bf x},u)=a_{\nu}(t,{\bf x})+ O(u),  \quad \Psi(t,{\bf x},u)= \Psi^- u^{\Delta^-} +  \Psi^+ u^{\Delta^+}.
\end{equation}
where
\begin{equation}
\Delta^\pm = \frac{4 \pm \sqrt{16 + 4 m^2}}{2}.
\end{equation}
We take $m^2=-3$, 
$a_\nu$ defines a background gauge field for the $U(1)$ current $j^{\nu}$ of the dual theory, with $\Psi^-$
an external source for the condensate $\Psi^+ = \psi$, where $\psi$ is the operator dual to the scalar
field.
Due to the  scaling symmetry of equations of motion, the  temperature is proportional $1/\mu$ which means we can
set $u_h=1$, increasing $\mu$ we can effectively reduce the temperature to induce a superfluid phase transition.

The superfluid phase we are interested is the spontaneous broken phase with finite 
chemical potential, zero external superfluid velocity, 
and the external sources $\Psi^-$ has to be set to zero on the boundary,
\begin{equation}
a_t(t,{\bf x})=\mu, \quad \textbf{a}=0,  \quad  \Psi^-=0.
\end{equation}
the expectation value of the superfluid condensation is determined by the subleading asymptotics of $\Psi$
\begin{equation}
\langle \psi(t,{\bf x}) \rangle= \lim_{u\rightarrow0} \partial_u^3 \Psi(t,{\bf x},u).
\end{equation}
Such a theory has a $U(1)$ symmetry broken solution admit lowest free energy  when the chemical
potential is above a critical value $\mu_c =4.16$. At the boundary velocity $\textbf{a}$ is set to be zero.
Then the gauge invariant velocity of boundary superfluid $\textbf{v}= \nabla \theta-\textbf{a}=\nabla \theta$,
where $\theta$ is the order parameter phase.
We choose $\mu=5$ which
corresponds to a superfluid state at temperature $T=0.83T_c$, for other temperatures similar
qualitatively similar results are obtained. 
{Initially we prepared a uniform superfluid state (shown in Fig. \ref{Figs1}) with zero superfluid velocity, which can be obtained by solving the 
equation of motion with the Newton-Raphson iteration method.
To introduce  the turbulence dynamics, we randomly imprint vortices to the uniform superfluid state by multiplying phase factor
$\prod^{N}_{i=1}\mathrm{exp}(i\phi_i)=\prod^{N}_{i=1}\mathrm{exp}(is_i \mathrm{arctan}[(y-y_i)/(x-x_i)])$ on each slice/plane of the global scalar field $\Psi(z_j,u)=|\Psi(z_j,u)| e^{i\phi(z_j)}$. The coordinates $(x_i,y_i,z_j)$ refer to the position of the i-th vortex on the plane $(z_j)$  of  $x-y$, where $j$ range from $1$ to grid size $50$, and $s_i= \pm 1$ corresponds to the winding number of the vortex. We repeat this step  on all the $x-y$ planes, $y-z$ planes and $z-x$ planes.}

\begin{figure}[t]
\centering
\includegraphics[trim=2.0cm 9.4cm 4.9cm 9.9cm, clip=true, scale=1, angle=0]{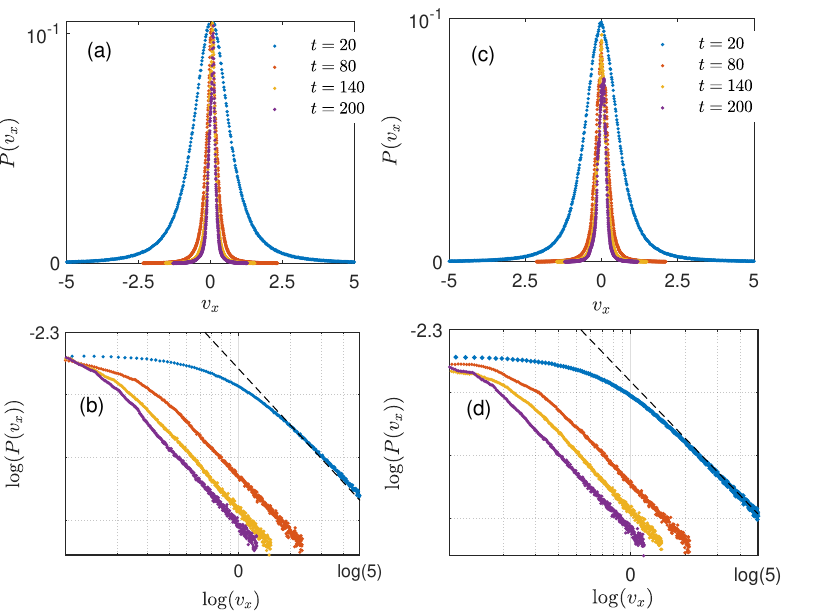}
\caption{{Velocity statistics and it's Log-Log plot for qusiclassic decay (a-b) and ultraquantum decay(c-d) at
different time,} the dashed line in the Log-Log plots is the power $-3$ line, clearly in both cases the velocity statistics show the same non-Gaussian properties different
from classic turbulence.}\label{Fig6}
\end{figure}

\begin{figure}[t]
\centering
\includegraphics[scale=1, angle=0]{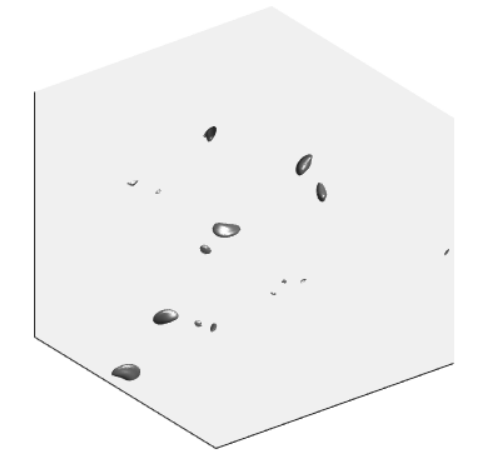}
\caption{$dE/dt$ through the horizon at time $t = 140$ for the quasi-classical decay case. The flux is zero nearly everywhere except at the locations of vortex  lines decay events.}\label{Fig5}
\end{figure}

\section{Method of generating turbulence and numerical details of solving PEDs }
\label{appendixB}
The phase configuration $\theta(u,\textbf{x})$ of $\Psi$ for a 2D vortex with winding number $W=\pm 1$ is known,
which is independent of AdS radial  coordinate $u$ since the vortex core is
stretching from the boundary to the horizon as a flux tube.
The initial velocity $\textbf{v}=\nabla \theta$ given by the
initial random placed 2D vortices with number $N$ in every slice (2D plane) is random; hence, the initial state
is dynamically unstable and soon produces homogeneous
and isotropic turbulence with many quantized
vortex loops. The number of the vortex loops is approximately
proportional to $N$.

In order to  solve the highly non-linear  bulk
equations of motion (PDEs),
we adapt the pseudospectral methods in the spatial directions,
all fields in a basis of 31 Chebyshev
polynomials in the radial direction and 181 plane waves in each boundary spatial direction.
In the time revolution the fourth order Rungle-Kutta method was used, the time step is
$\delta t= 0.05$. {In Fig. \ref{Figs2} three samples of $|\Psi(u)|$ are shown in a  turbulent moment, corresponding to 
three different positions from that far away from a vortex line 
to a position close to the vortex line core. Compared to the initial 
uniform static case shown in Fig. \ref{Figs1}, the superfluid density is suppressed due to the existence of supercurrent around a vortex line core.}

{\section{Decay of vortex lines at later times}}

\begin{figure}[t]
\centering
\includegraphics[scale=1, angle=0]{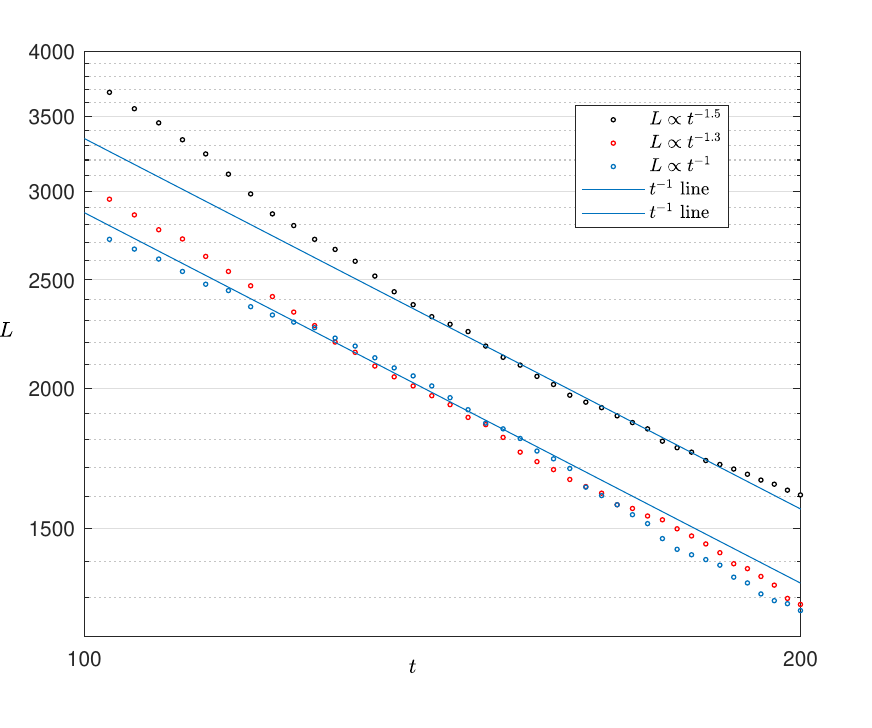}
\caption{Compare decay of vortex lines in late time for different initial vortex line densities.}\label{latertime}
\end{figure}

{Here we show the details of vortex line decay dynamics at late times for different initial vortex line densities (see Fig. \ref{latertime}). Though all the decay dynamics approach $t^{-1}$ near $t\sim1 70$, it can be found that the deviation from $t^{-1}$
near $t=200$ always happens, not only for the quasi-classical case but also for the
ultra-quantum decay case. {In our opinion, the late time regime may not be defined as a  turbulent state
because the vortex lines are very few, and so Vinen's decay equation
should not be applied since it is a statistical results for many vortex lines. The decay dynamics with very few vortex lines seems to be complex and depending on the concrete spatial configurations.}}

\end{appendix}

\end{document}